\documentclass[journal=jacsat,manuscript=article]{achemso}

\usepackage{chemformula} % Formula subscripts using \ch{}
\usepackage[version=3]{mhchem} % Formula subscripts using \ce{}
\usepackage[T1]{fontenc} % Use modern font encodings
\usepackage{soul}
\author{Dorothy Gogoi}
\affiliation[2]{School of Physical Sciences, Jawaharlal Nehru University, New Delhi-110067, India.}
\author{Avinash Chauhan}
\affiliation[1]{Department of Physics, Indian Institute of Technology (BHU), Varanasi, Uttar Pradesh-221005, India.}
\author{Sanjay Puri}
\affiliation[2]{School of Physical Sciences, Jawaharlal Nehru University, New Delhi-110067, India.}
\email{purijnu@gmail.com}
\author{Awaneesh Singh}
\email{awaneesh11@gmail.com}
\affiliation[1]{Department of Physics, Indian Institute of Technology (BHU), Varanasi, Uttar Pradesh-221005, India.}

\SectionsOn
\SectionNumbersOn
\title[]
{Segregation Kinetics of Miktoarm Star Polymers: A Dissipative Particle Dynamics Study}

\begin{document}

%\begin{tocentry}
%\begin{center}
%\includegraphics[width=0.9\textwidth]{TOC.png}
%\end{center}
%\end{tocentry}

\begin{abstract}
We study the phase separation kinetics of miktoarm star polymer (MSP) melts and blends with diverse architectures using dissipative particle dynamics simulations. Our study focuses on symmetric and asymmetric miktoarm star polymer (SMSP/AMSP) mixtures based on arm composition and number. For a fixed MSP chain size, the characteristic microphase-separated domains initially show diffusive growth with a growth exponent $\phi \sim 1/3$ for both melts that gradually crossover to saturation at late times. The simulation results demonstrate that the evolution morphology of SMSP melts exhibits perfect dynamic scaling with varying arm numbers; the time scale follows a power-law decay with an exponent $\theta \simeq 1$ as the number of arms increases. The structural constraints on AMSP melts cause the domain growth rate to decrease as the number of one type of arms increases while their length remains fixed. This increase in the number of arms for AMSP corresponds to increased off-criticality. The saturation length in AMSP follows a power law increase with an exponent $\lambda \simeq 2/3$ as off-criticality decreases. Additionally, macrophase separation kinetics in SMSP/AMSP blends show a transition from viscous ($\phi \sim 1$) to inertial ($\phi \sim 2/3$) hydrodynamic growth regimes at late times; this exhibits the same dynamical universality class as linear polymer blends, with slight deviations at early stages.
\end{abstract}

\section{Introduction}
\label{Intro}
There has been significant interest in studying phase separation (PS) kinetics in polymeric fluids for decades due to their vast industrial \cite{asenjo2012aqueous, matyjaszewski1999synthesis} and technological applications \cite{xue2012phase, hawker1995architectural}. When a multicomponent fluid is quenched from a homogeneous phase to a point deep inside the coexistence curve, it becomes thermodynamically unstable \cite{puri2009, bray1994, dattagupta2004}. This \emph{far-from-equilibrium} system approaches equilibrium by evolving into domains enriched in either of the fluid components \cite{puri2009, bray1994}. The characteristic length scale of evolving domains [$L(t)$, where $t$ is the time], typically exhibits power-law dependence: $L(t) \sim t^{\phi}$, where the growth exponent $\phi$ depends on the transport mechanism that drives the segregation \cite{puri2009, bray1994, dattagupta2004}. 

Domain coarsening is a well-established scaling phenomenon \cite{bray1994, puri2009, dattagupta2004}, as manifested by the two-point equal-time \emph{correlation function}, $C(\vec{r},t)$ which obeys $C(r,t) = g[r/L(t)]$. Here, $C(r,t)$ is the spherically-averaged correlation function, $\vec{r}=\vec{r}_2-\vec{r}_1$, $r = |\Vec{r}|$ and $g(x)$ is a scaling function which is independent of time \cite{bray1994, puri2009}. Diffusive growth ($\phi= 1/3$ in $d \geq 2$) is the only expected growth mechanism \cite{bray1994, puri2009} in phase-separating binary alloys. However, for fluid mixtures (e.g., oil-water, polymer blends), one expects larger growth exponents resulting from the advective transport of their components. For example, in $d=3$ mixtures, $\phi = 1/3$ for $L(t)\ll \left(D\eta\right)^{1/2}$, $\phi = 1$ for $\left(D\eta\right)^{1/2} \ll L(t)\ll L_{in}$, and $\phi = 2/3$ for $L(t)\gg L_{in}$ \cite{bray1994, puri2009}. Here, $\phi = 1$ and $2/3$ are referred to as the viscous and inertial hydrodynamic growth exponents respectively; $L_{in}\simeq\eta^2/(\rho \sigma)$ is the inertial length, and $D$, $\eta$, $\rho$, and $\sigma$ denote the diffusion coefficient, viscosity, number density, and interfacial tension, respectively. Unlike polymer blends, linear block copolymer (BCP) ($A_nB_m$) melts undergo microphase separation (micro-PS) while maintaining chemical connectivity between incompatible $A$ and $B$ subchains with degrees of polymerization $n$ and $m$, resulting in microscale morphologies \cite{leibler1980theory, bates1991polymer, kim2010functional, groot1998dynamic, groot1999role, chakrabarti1989microphase}. At early times, the BCP segregation is similar to the usual spinodal decomposition in fluids or polymer blends ($\phi \sim 1/3$) \cite{singh2015kinetics, liu1989dynamics, groot1998dynamic, groot1999role}. However, at late times, the BCP evolution freezes into a micro-scale morphology dictated by the $n:m$ ratio, e.g., lamellar, cylindrical, droplet \cite{bates1991polymer, liu1989dynamics}.

The miktoarm star polymers (MSPs) have unique topologies that have garnered significant attention in recent times \cite{BEZIAU2016, wei2005synthesis}. They are branched polymers of a distinct class featuring multiple arms that meet at a central junction \cite{wei2005synthesis, yuan2005synthesis} with varying molecular, structural, and compositional properties. In contrast to conventional linear BCPs, MSPs can self-assemble into multi-compartment micelles, worm-like micelles, and other unique designs \cite{mountrichas2005micelles, li2004multicompartment, saito2010multicompartment}. These polymers are investigated extensively to explore their applications in diverse fields, including drug encapsulation and delivery systems \cite{wu2015star, lotocki2020miktoarm}, stimuli-responsive materials \cite{khanna2010miktoarm}, and fabricating nanostructured thin films \cite{ren2016star, workineh2020tuning}. The micro-PS of MSPs is used in creating different Janus structures such as Janus micelles, cylinders, and discs \cite{Erhardt2001, Erhardt2003}. Experimental studies have examined the synergistic effect of chain length and solvent conditions on the self-assembled structures of various MSPs \cite{Tsitsil1995, Tsitsil2006, floudas1997microphase}.

It has been observed that the self-assembly of MSPs is more challenging than BCPs due to the increasing number of constraints \cite{grason2004interfaces}. A dissipative particle dynamics (DPD) simulation \cite{espanol1995statistical, groot1997dissipative} study involving MSP melt, where an MSP molecule consists of one $A$-type arm and two $B$-type arms \cite{huang2007microphase}, has analyzed the micro-PS behavior and its equilibrium conformation; the study reported that a small volume fraction of $B$ block forms only a tube-like phase, which is often observed in experiments. Recent coarse-grained simulation studies have demonstrated that the phase separation of MSPs in bulk can be substantially influenced and customized by varying the number and length of arms \cite{li2014coarse, tan2015dissipative}. Most of these studies are conducted in solvent medium with smaller system sizes, focusing on studying the final equilibrium morphology. Therefore, it is crucial to investigate and devise effective methods to study the dynamical properties of various segregating MSP systems, which have not been explored considerably. Thus, this work extensively studied the phase separation kinetics of various symmetric and asymmetric MSP (SMSP/AMSP) melts and blends, focusing mainly on the evolution morphologies and growth laws using a generic DPD framework. The main objective of this study is to gain a numerical understanding of phase separation kinetics in complex systems like MSPs, where theoretical calculations are currently complicated.
\begin{figure}[tb]
\centering
  \includegraphics[width=0.78\textwidth]{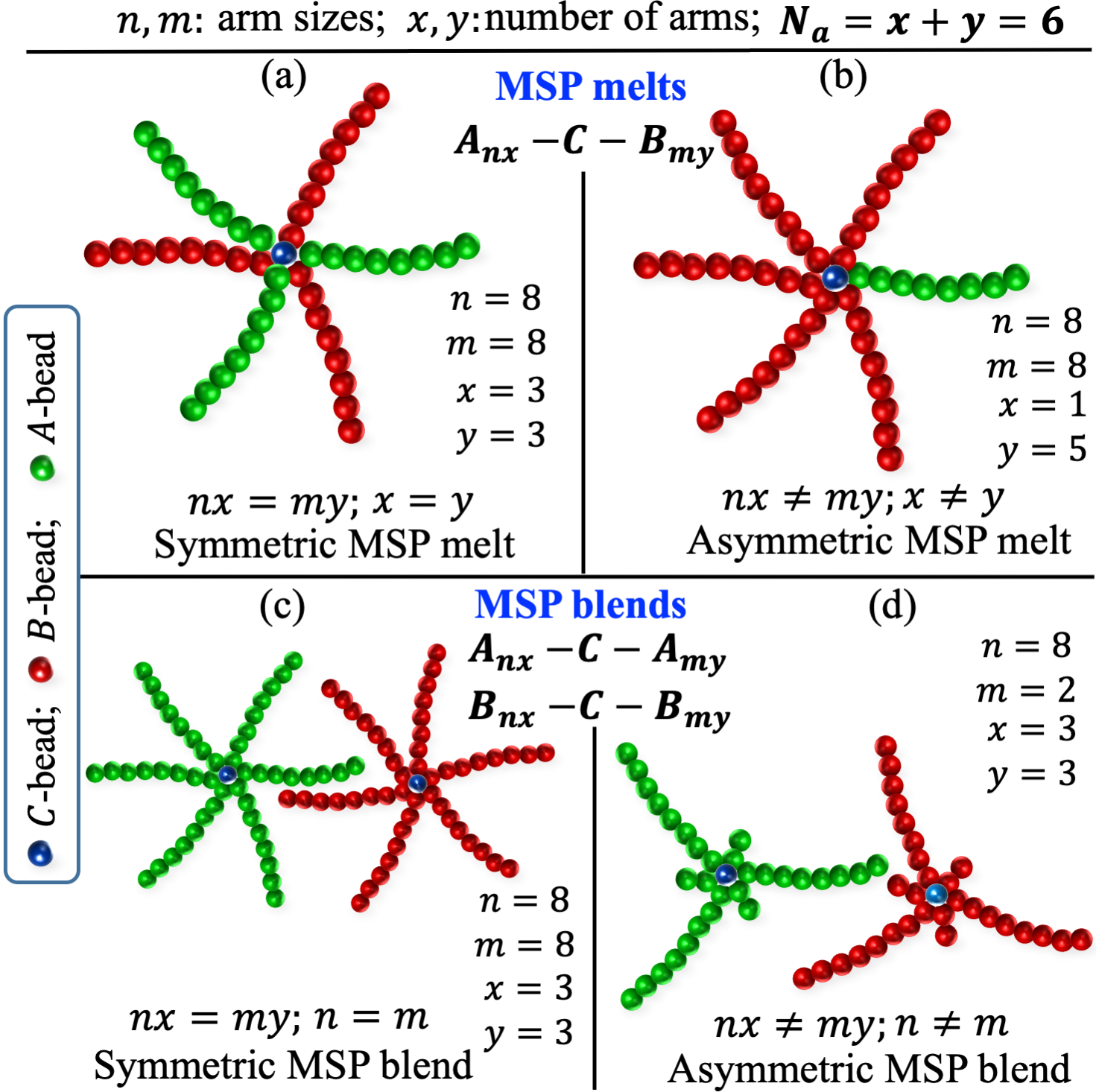}
  \caption{Schematic representation of 4 distinct configurations of $6$-arm MSPs. Here, $A$ beads are shown in green, $B$ in red, and the central $C$ in blue. Further details about their architecture can be found in the text.}
  \label{Fig1}
\end{figure}

In our study, we examine both MSP melt and MSP blend systems. An MSP melt is a mixture consisting solely of a single type of MSP molecule. In contrast, an MSP blend denotes a mixture of two types of MSP molecules. Our study first considers MSP melts of type $A_{nx}-C-B_{my}$. Each MSP consists of incompatible $A$ and $B$ arms; $x$ and $y$ are the number of arms, and $n$ and $m$ are the degree of polymerization of $A$ and $B$ arms, respectively. The arms are connected to a central bead $C$, forming an MSP molecule with $N_a=x+y$ arms and $N_b=nx+my+1$ beads. We characterize SMSP melts by $n=m$ and $x=y$, and AMSP melts by $n=m$ with $x \neq y$, where $x=1$ is fixed. The asymmetry in arm compositions characterizes the AMSPs we study. There is a corresponding asymmetry in volume fractions of $A$ and $B$ phases, and the composition changes from critical to off-critical melts by variation in $y$. The schematic diagrams in Fig.~\ref{Fig1}(a-b) illustrate a specific case of SMSP and AMSP molecules in melts. Note that $N_a=2$ denotes a BCP molecule. This allows us to investigate and compare the phase separation kinetics of multiarm MSP ($N_a>2$) melts with the well-studied BCP melt. Second, we consider the blends of $A_{nx}-C-A_{my}$ and $B_{nx}-C-B_{my}$ type of MSPs. We choose $n=m$ and $x=y$ for the SMSP molecule, and $n\neq m$ (asymmetry in arm molecular weight or degree of polymerization) and $x=y$ for the AMSP molecule in the blend (see Fig.~\ref{Fig1}(c) and \ref{Fig1}(d)). Notably, $N_a=2$ resembles a conventional linear homopolymer. We focus on how systematically increasing $N_a$ ($A$ and $B$ arms of MSPs) influences their phase separation kinetics in both cases. The number of $C$ beads is significantly smaller than $A$ and $B$. Therefore, we effectively treat our systems as binary melts or blends. The color scheme for MSP beads is consistent across all configurations, with green, red, and blue representing $A$, $B$, and $C$ beads, respectively. Further details are provided in the following sections.

%%%%%%%%%%%%%%%%%%%%%%%%%%%
\section{Methodology}
\label{method}
The DPD simulation technique is used to model the various systems \cite{groot1997dissipative}. This is a powerful and robust numerical approach for studying the dynamics of diverse complex systems at the mesoscopic scale \cite{groot1997dissipative, espanol1995statistical, espanol2017perspective}. In DPD, the elementary unit is a coarse-grained soft bead representing a small fluid region containing a group of particles or molecules. This makes it a more viable numerical tool for simulating the system over longer length and time scales than traditional molecular dynamics (MD) simulation technique \cite{groot2006local, nikunen2003would, nikunen2007reptational}. The time evolution of the system involves integrating Newton's equation of motion to update the position and velocity of each bead:
\begin{equation}
\label{Newton}
\vec{f}_i(t) = m_i\frac{d\vec{v}_i}{dt},
\end{equation}
where $\vec{v}_i = d\vec{r}_i/dt$, $m_i$, and $\vec{r}_i$ correspond to the velocity, mass, and position vector of the $i^{th}$ bead. The force, $\vec{f}_i(t)$, on the $i^{th}$ bead exerted by all other $j$ beads within a cutoff distance $r_c$ is composed of three pairwise additive forces as follows:
\begin{equation}
\label{Forces}
\vec{f}_i(t)= \sum_{j\neq i}{ \left(\vec{F}_{ij}^C+\vec{F}_{ij}^D +\vec{F}_{ij}^R \right)}.
\end{equation}
Here, $\vec{F}_{ij}^C$, $\vec{F}_{ij}^D$, and $\vec{F}_{ij}^R$ are the conservative, dissipative, and random forces, respectively, defined as \cite{espanol1995statistical, groot1997dissipative}:
\begin{eqnarray}
\label{Force}
\vec{F}_{ij}^C &=& a_{ij}\left(1-{r_{ij}}/{r_c}\right)\hat{r}_{ij},\\
\vec{F}_{ij}^D &=& -\gamma_D \omega_D(r_{ij})(\hat{r}_{ij} \cdot \vec{v}_{ij})\hat{r}_{ij},\\
\vec{F}_{ij}^R &=& \sigma_R\omega_R(r_{ij}) \xi_{ij} \hat{r}_{ij}.
\end{eqnarray}

The conservative force is a soft repulsive interaction, where $a_{ij}$ signifies the maximum repulsion between the $i^{th}$ and $j^{th}$ beads; its magnitude is determined by the type of beads involved. This interaction is linear up to a cutoff distance $r_c$ and zero otherwise. It characterizes the pairwise potential between the $i^{th}$ and $j^{th}$ beads, acting along the line connecting their centers. In this context, $\vec{r}_{ij}=\vec{r}_{i}-\vec{r}_{j}$ and $\vec{v}_{ij}=\vec{v}_{i}-\vec{v}_{j}$ represent the relative position and velocity, respectively; $r_{ij}=\left| \vec{r}_{ij}\right|$, and the unit vector $\hat{r}_{ij}=\vec{r}_{ij}/r_{ij}$ determines the direction of the force.

The interplay between dissipative and random forces serves as a thermostat, with the dissipative force acting as a heat sink and the random force acting as a heat source. The Gaussian random variable, $\xi_{ij}$, is characterized by a zero mean and unit variance: 
\begin{equation}\label{Gauss}
\begin{aligned}
&\langle \xi_{ij}(t)\rangle=0,  \\
&\langle \xi_{ij}(t) \xi_{kl}(t^{\prime})\rangle=(\delta_{ik}\delta_{jl} + \delta_{il}\delta_{jk} )\delta(t-t^{\prime}).
\end{aligned}
\end{equation}
Here, $\langle \cdots \rangle$ denotes the ensemble average. The parameters $\gamma_D$ and $\sigma_R$ represent the strengths of the dissipative and random forces, respectively, referred to as the friction coefficient and noise strength. They are linked through the fluctuation-dissipation theorem to maintain the correct canonical state of the system \cite{espanol1995statistical, groot1997dissipative}:
\begin{equation}
\label{fdt}
\sigma_R^2=2\gamma_{D} k_BT/m,
\end{equation}
where, $T$ is the system temperature, and $k_B$ is the Boltzmann constant. The weight functions $\omega_D$ and $\omega_R$ define the interaction range for the dissipative and random forces within a cutoff distance $r_c$. They are governed by the typical relation \cite{espanol1995statistical,groot1997dissipative}:
\begin{equation}\label{omega}
\omega^D(r_{ij})=\left[\omega^R(r_{ij})\right]^2=\left(1-r_{ij}/r_c\right)^2.
\end{equation}
These forces conserve local momentum and thus preserve correct hydrodynamic behavior even in small systems consisting of a few hundred beads. Therefore, DPD has the advantage of naturally incorporating flow fields and preserving the hydrodynamics in the system \cite{groot1997dissipative}. 

The cutoff distance $r_c$ and $k_{B}T$ represent the characteristic length and energy scales of the system, respectively. Each DPD bead is assigned an identical mass $m_i=m$. All our results are presented in reduced DPD units where $r_c$, $m$, and $k_BT$ are set to unity \cite{groot1997dissipative}. The number density is fixed at $\rho = 3r_c^{-3}$ \cite{groot1997dissipative}, which keeps the system far from the gas-liquid transition during simulation. We integrate the equation of motion using the modified \textit{velocity-Verlet} algorithm \cite{plimpton1995fast} with a time step of $\Delta t=0.02\tau$. Here, the characteristic time scale $\tau = {(m{r^2}_c/k_BT)}^{1/2} = 1.0$ (in reduced DPD units). We set $\gamma_D = 4.5$ (in reduced DPD units), a commonly used value in DPD simulations, as it strikes a good balance between accurately capturing hydrodynamic behavior and maintaining computational efficiency across various systems and conditions \cite{hashimoto1986late, singh2016tailoring, singh2017photo, chen2017living}. This value ensures that the dissipative force is strong enough to mimic hydrodynamic interactions while ensuring numerical stability within the selected time steps \cite{hashimoto1986late, groot1997dissipative}. However, the choice of $\gamma_D$ may vary depending on the specific system under study and the desired level of detail required in the simulation \cite{espanol1995statistical, groot1997dissipative}.

In our simulation, ten water molecules are coarse-grained to form a DPD bead, representing a volume of $300$ \AA$^{3}$ for the water mass density, $\rho_m = 1 \; g/cm^{3}$ \cite{singh2023additive, singh2023amph}. This choice, with $\rho = 3r_{c}^{-3}$, yields $r_c \simeq 0.97 \; nm$ and the characteristic mass, $m = 180 \; Da$ \cite{groot1997dissipative, chen2017living, singh2023additive}. Utilizing these characteristic length, mass, and energy $(k_BT)$ at $T=297 \; K$, the estimated dimensional unit of time is $\tau \simeq 8.3 \; ps$ \cite{groot1997dissipative, singh2023additive, singh2023amph}, which represents the intrinsic time scale that accelerates the system's dynamics, owing to the softcore potential \cite{symeonidis2006schmidt, junghans2008transport}. To establish a more relevant time scale, one can correlate the diffusion coefficient of the melt from DPD simulations with its corresponding experimental value \cite{Kremer1990, Wim2018, Yong2013}. For example, the typical diffusion coefficient of a melt in DPD simulations is $D_{sim} \sim 10^{-2} \; nm^2/\tau$ \cite{Yong2013}, while the experimental value is $D_{exp} \sim 10^{-11} \; m^2/s$ \cite{nikoobakht2003preparation, Janez2020, hsu2016static}. Therefore, a more suitable simulation DPD timescale for the system is $\tau \sim 1.0 \; ns$ \cite{Wim2018, junghans2008transport, chen2017living, singh2023additive}.

In DPD, the interaction parameter, $a_{ij}$, determines the repulsive force strength among beads, thereby influencing the structural and dynamical characteristics of the system \cite{groot1997dissipative}. In systems with multiple bead types, the relative values of $a_{ij}$ influence the phase separation kinetics of the system \cite{groot1998dynamic, singh2018role}. For instance, in a binary ($AB$) mixture, higher $a_{AB}$ compared to $a_{AA}$ and $a_{BB}$ can induce phase separation between $A$ and $B$ beads \cite{groot1997dissipative, groot1999role, singh2021photo}. Precisely adjusting these parameters is imperative for accurately simulating complex fluid systems \cite{groot1997dissipative}. We set $a_{ii} = 25$ (in the units of $k_BT/r_c$) between beads of the same type, such as $a_{AA}= a_{BB} = a_{CC} = 25$ \cite{groot1997dissipative, singh2016tailoring, singh2017photo}. For simplicity, the $C$-type bead is considered compatible with other beads, i.e., $a_{AC} = a_{BC} = 25$. For the interaction between incompatible $A$ and $B$ beads, $a_{ij} = a_{ii} + 3.27 \chi_{ij}$ \cite{groot1997dissipative, singh2023amph, chen2017living, singh2023additive}. Here, $\chi_{ij}$ represents the interaction parameter between polymer beads in the Flory-Huggins lattice model. We choose $\chi_{ij}\simeq 10.7$ (in reduced DPD units), resulting in $a_{AB} = 60$ \cite{groot1998dynamic, groot1997dissipative, singh2023additive, singh2023shearbcp, singh2023amph}. Shortly, we observe that our chosen parameter values facilitate phase separation in the system \cite{groot1999role, groot1998dynamic, singh2023shearbcp, singh2023additive, singh2023amph}. 

We employ the \emph{bead-spring model} to simulate polymer chains \cite{junghans2008transport, hsu2016static, singh2016tailoring, singh2017photo, nikunen2007reptational, karatrantos2013topological}, where beads within each chain are connected by a harmonic bond potential \cite{junghans2008transport, hsu2016static}: 
\begin{equation}\label{bp}
E_b = \dfrac{\kappa_b}{2}\left(r - r_0\right)^2.
\end{equation}
The chain stiffness is controlled by the angle potential, expressed as
\begin{equation}\label{ap}
E_{a}=\frac{\kappa_a}{2}\left(\cos \theta - \cos \theta_0\right)^2.
\end{equation}
The rigidity parameters, $\kappa_b$, and $\kappa_a$ define the strength of these potentials. In this context, $r_0 = 0.5$ signifies the equilibrium bond length, $\theta$ denotes the angle between sequential bonds along the chain, and $\theta_0=180^{\circ}$ represents the equilibrium value of the angle \cite{hsu2016static}. Note that the angle potential penalizes deviations in $\theta$ from $\theta_0$, which prevents excessive bending or stretching of polymer chains and, thus, maintains realistic conformations during simulation. Therefore, setting $\theta_0 = 180^{\circ}$ is a practical strategy for simulating generic linear polymers, as it effectively preserves chain linearity and stabilizes bond angles \cite{junghans2008transport, hsu2016static}. Our simulation model is not tailored to any specific polymeric system. Therefore, we use commonly employed rigidity parameters: $\kappa_b=128$ and $\kappa_a=5$ (in reduced DPD units) for all bonds and angles of flexible polymer chains \cite{singh2016tailoring, singh2017photo, nikunen2007reptational, karatrantos2013topological}.

It is important to note that the softcore interaction among beads can lead to undesired bond crossings. To address this issue, we implement the modified \emph{segmental repulsive potential} (mSRP) \cite{sirk2012enhanced}. This potential acts specifically on bond pairs, treating individual bonds as \emph{fictitious beads}, which interact through a soft repulsive interaction given by,
\begin{equation}\label{srp}
\vec{F}_{ij}^S = \kappa_S\left(1 - \frac{r_{ij}}{r_c^S}\right) \hat{r}_{ij},  \text{ for $r_{ij} \le r_c^S$}.
\end{equation}
These fictitious beads are specified at the beginning of the simulation, serving as markers that denote the bond positions. This approach facilitates the creation of neighbor lists and enables the computation of pairwise interactions akin to that for actual beads. Such an approach impedes the interpenetration of bonds when subjected to a soft, non-bonded potential amid beads in DPD polymer chains. We set the coefficient $\kappa_S=100$ and the cutoff distance $r_c^S=0.8$ \cite{singh2023shearbcp, singh2023additive, singh2023amph, olga2022deg, olga2023bb}. The equations of motion are integrated using the LAMMPS simulation package \cite{LAMMPS} with the mSRP code \cite{sirk2012enhanced}.

%%%%%%%%%%%%%%%%%%%%%%%%%%%%%%%%%%%%%%%%%%%%%
\section{Results and discussions}
%%%%%%%%%%%%%%%%%%%%%%%%%%%%%%%%%%%%%%%%%%%%%
\subsection{Morphology characterization}
\label{morphology}
%%%%%%%%%%%%%%%%%%%%%%%%%%%
To characterize evolution morphologies, we calculate the radial distribution function (RDF) described as: 
\begin{equation}
g_{\alpha\beta}(r) = \rho_{\alpha\beta}(r)/\Bar{\rho}_{\beta}.
\label{gr1}
\end{equation}
Here, $\rho_{\alpha\beta}(r)$ signifies the local density of $\beta$-type beads surrounding the reference $\alpha$-type bead, and $\Bar{\rho}_{\beta}$ denotes the average total density of $\beta$-type beads. Further, we utilize the two-point equal-time correlation function $C(\vec{r},t)$ \cite{bray1994, puri2009} defined by:
\begin{equation}
C\left(\vec{r}, t\right) = \left\langle S\left(\vec{r}_1, t \right) S\left(\vec{r}_2, t\right)\right\rangle  -  \left\langle S\left(\vec{r}_1, t \right)\right\rangle \left\langle S\left(\vec{r}_2, t \right)\right\rangle,
\label{cr1}
\end{equation}
where $ S\left(\vec{r}_1, t \right)$ and $ S\left(\vec{r}_2, t \right)$ denote the order parameter at two discrete sites $\vec{r}_1$ and $\vec{r}_2$ at a given time $t$, while $\vec{r}=\vec{r}_2 - \vec{r}_1$ signifies the distance between these sites. The angular brackets denote the ensemble average. The order parameter, $S\left(\vec{r}, t \right)$, is computed using a coarse-grained technique. The simulation box is divided into non-overlapping unit-size boxes, and the continuum fluid is mapped within each box. The number of beads of each type $n_A$, $n_B$, and $n_C$ is counted in each box. We assign order parameter values $S=+1$ and $-1$ when the maximum number of beads in a unit box is of type $A$ and $B$, respectively \cite{singh2018role, singh2015kinetics}. For the maximum number of beads of type $C$, a value of $S=+1$ or $-1$ is randomly assigned with equal probability. In cases where, for instance, $n_A = n_B > n_C$, the order parameter value $+1$ or $-1$ is randomly assigned with equal probability. A similar approach is applied to other degeneracies.

The correlation function $C(\vec{r}, t)$ in our simulation is spherically averaged over five independent runs to improve statistics and denoted by $C(r,t)$ \cite{bray1994, puri2009}. There exist various definitions of the characteristic length scale $L(t)$ of the phase-separating systems, such as the distance at which the correlation function decays to a fraction of its maximum value or the inverse of the first moment of the structure factor \cite{bray1994, puri2009}. These definitions differ only by a constant multiplicative factor in the scaling regime \cite{bray1994, puri2009, oono1987computationally, oono1988study}. In this study, we adopt the definition of $L(t)$ as the first zero crossing of $C(r,t)$. Note that, for a conserved system, the correlation function $C(r,t)$ exhibits damped oscillations around zero, the value to which it asymptotically decays. 

For our simulation, we consider a cubic simulation box of size $L=64$ with periodic boundary conditions in all directions. At the onset of DPD simulation, we equilibrate the system up to $t=1.0\times 10^3$ at a high temperature of $T=5$. The time is then reset to $t=0$. Subsequently, we quench the system at $T=1$. As we will soon observe, this quench temperature is well below the critical temperature for inducing phase separation in MSP melts and blends. 

%%%%%%%%%%%%%%%%%%%%%%%%%%%%%%%%%%%%%%%%%%%%%%%%
\subsection{Symmetric MSP melt} 
\begin{figure}[tb]
\centering
  \includegraphics[width=0.7\textwidth]{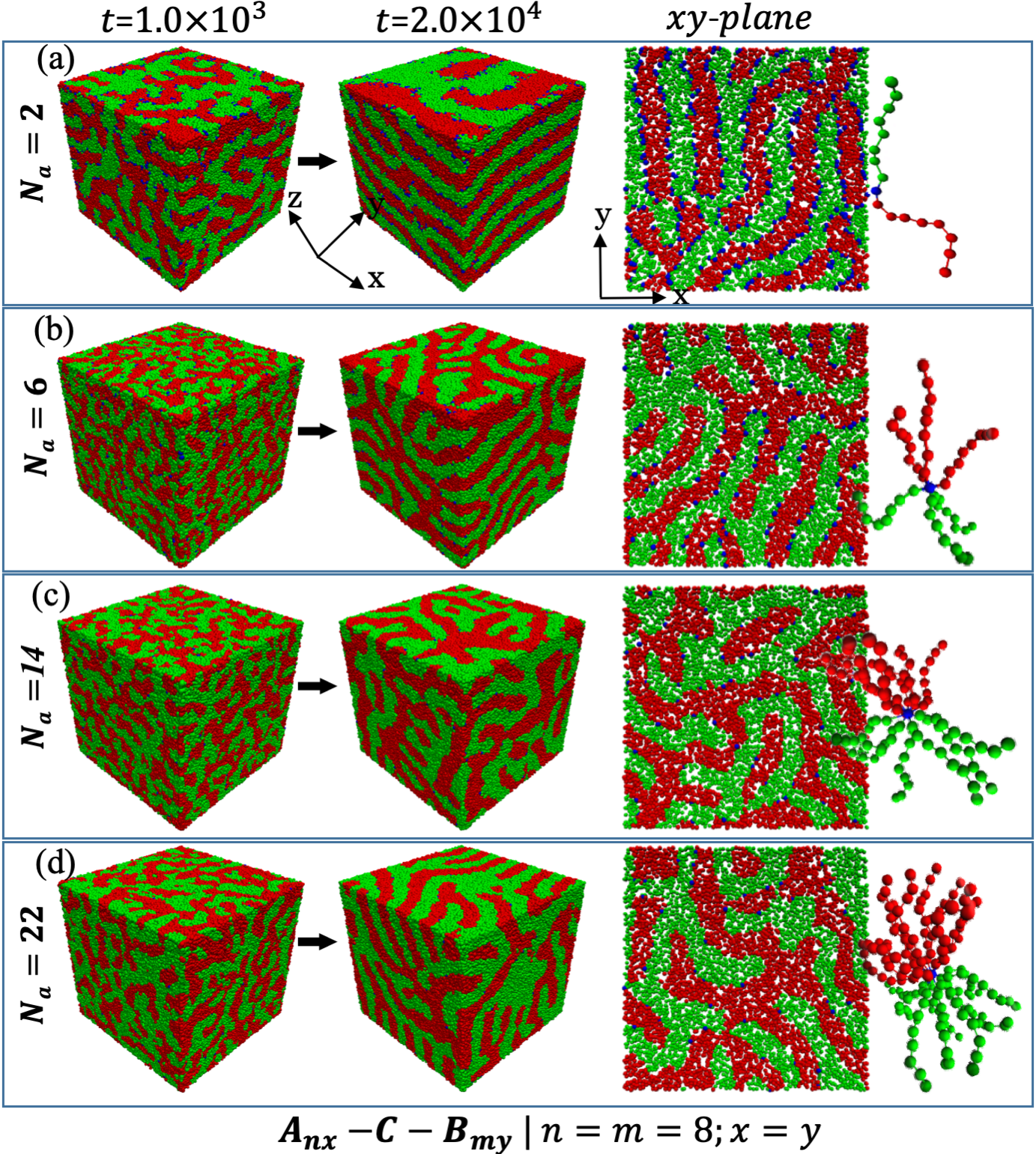}
  \caption{Evolution snapshots at $t=1.0\times10^3$ (first column) and $t=2.0\times10^4$ (second column) for $N_a=2$, $6$, $14$, and $22$ for an MSP melt of type $A_{nx}-C-B_{my}$ with $x=y$ and $n=m=8$ are shown. The third column displays $2d$ cross-sections ($xy$-plane at $z=32$) of the snapshots at $t=2.0\times10^4$. Images of individual MSP molecules are shown on the far right for each case.}
  \label{Fig2}
\end{figure}

%%%%%%%%%%%%%%%%%%%%%%%%%%%%%%%%%%%%%%%%%%%%%%%%
Our analysis begins with examining the evolution of SMSP $\left(A_{nx}-C-B_{my}\right)$ melts. The degree of polymerization for each arm is set to $n=m=8$, with an equal number of $A$ and $B$-type arms, i.e., $x=y$ per MSP. We study the phase separation kinetics as the number of arms, $N_a=x+y$, varies from $2$ to $22$. The $3d$ snapshots in Fig.~\ref{Fig2} exhibit the evolution morphology of MSP melts at $t=1.0\times10^3$ (first column) and $t=2.0\times10^4$ (second column) for (a) $N_a=2$, (b) $N_a=6$, (c) $N_a=14$, and (d) $N_a=22$. To clearly depict the morphology for each $N_a$, the third column presents the $xy$ cross-section (at $z=32$) corresponding to the morphologies in the second column. We observed that MSP with $N_a=2$ (which is a BCP) grows faster than the rest. However, the subsequent trend at higher $N_a$ is somewhat counterintuitive, as one may naively expect domain growth to slow down with an increase in $N_a$. Interestingly, MSP with $N_{a}=14$ and $N_a=22$ seem to form slightly larger domains than the $N_a=6$ case. After an initial transient regime, the local rearrangement of $A$ and $B$ arms leads to the formation of a lamellar structure at later times in all cases, which is typical of symmetric BCP melts. The images on the far right depict the microscopic structure of an individual MSP molecule in the melt. Clearly, all the like arms cluster on one side, suggesting a topology to that of the BCP. It is important to note that MSPs possess significantly different and complex molecular structures compared to much simpler linear BCPs. Therefore, their structural properties and dynamic scaling behaviors cannot be directly compared.

\begin{figure}[tb]
\centering
  \includegraphics[width=0.75\textwidth]{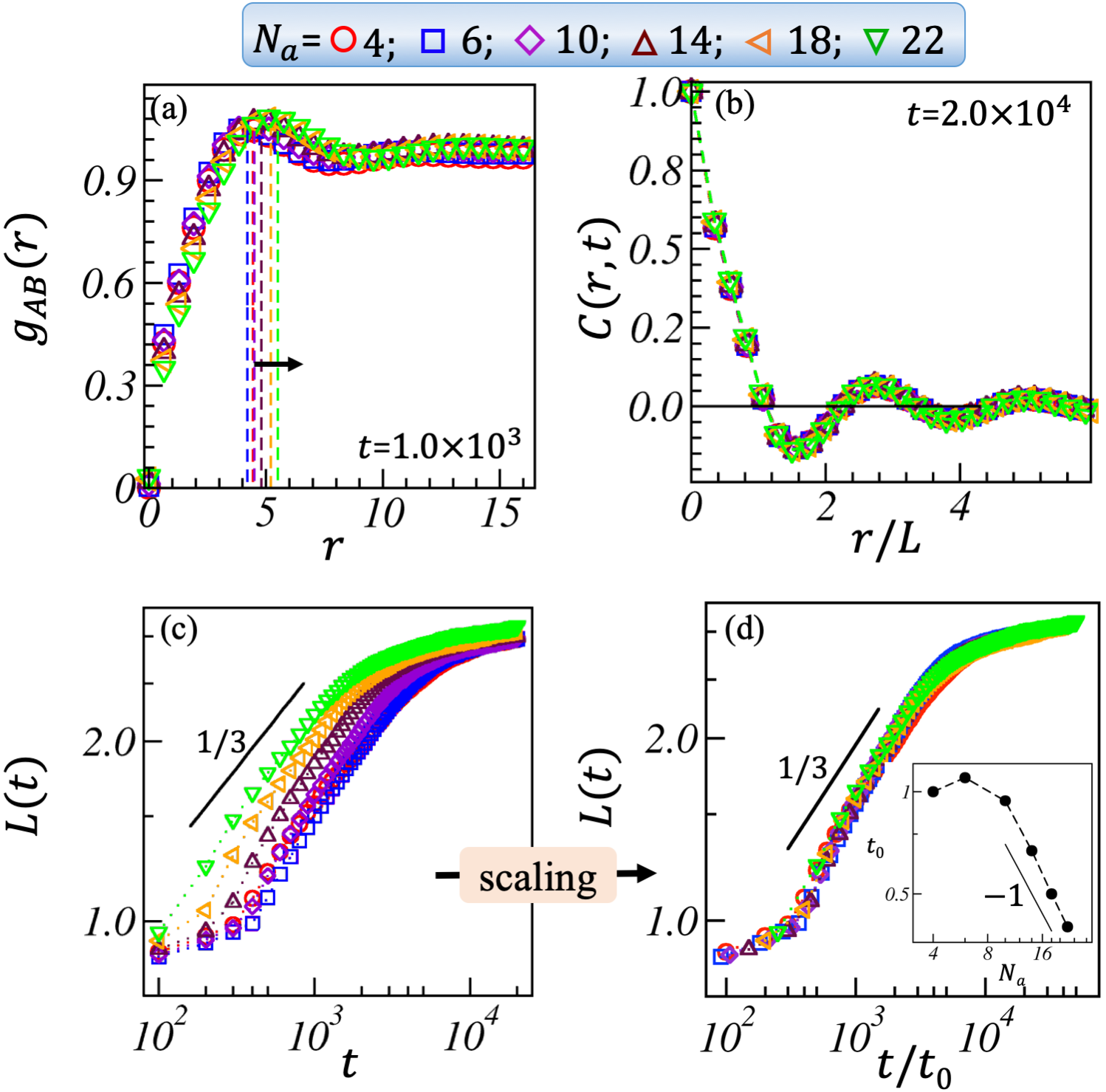}
  \caption{(a) RDF plots, $g_{AB}$ vs. $r$ for $N_{a}=4$ (red), $6$ (blue), $10$ (purple), $14$ (maroon), $18$ (orange), and $22$ (green) at $t=1.0\times 10^3$. (b) The plot of $C(r,t)$ vs. $r/L$ at $t=2.0\times 10^4$ for the same $N_{a}$ vales as in (a). (c) $L(t)$ vs. $t$ for the same $N_{a}$ values as in (a), plotted on a log-log scale. A black solid line of slope $1/3$ indicates the early diffusive growth regime. (d) Scaling of length scale $L(t)$ is obtained by scaling the time axis by a time factor $t_0$. The inset shows the plot of $t_0$ vs $N_a$.}
  \label{Fig3}
\end{figure}

We validated our observations in Fig.~\ref{Fig2} through a detailed study of morphology characterization using the tools defined in section \ref{morphology}. Figure~\ref{Fig3}(a) shows the radial distribution function (RDF), $g_{AB}(r)$ vs. $r$, for $N_a=4$, $6$, $10$, $14$, $18$, and $22$ at $t=1.0\times10^3$. The same RDF peak strength implies a similar distribution of $B$ beads around $A$ in each case. However, a noticeable trend is noted in the peak positions of the RDFs. The RDF peak for $N_a=4$ is located at a slightly higher $r$ compared to $N_a=6$, indicating a larger average cluster size \cite{Jack2017}. Note that the diffusion coefficient varies as $D \sim N_{b}^{-\nu}$, where $N_{b}$ is the number of beads per MSP and $\nu\sim 1/2$ in melts \cite{de1979scaling, jiang2007hydrodynamic}. Therefore, increasing $N_b$ of an MSP molecule leads to smaller $D$. Typically, the size of MSP molecule (measured by the radius of gyration) increases as $R_g \sim N_b^{\mu}$, with $\mu \sim 1/2$ for $N_a \leq 6$, where MSP molecules behave like linear polymer chains in the melt \cite{Jack2017}.

Upon increasing the number of arms further ($N_a\geq 6$), the RDF peak positions gradually shift to higher $r$. This behavior can be attributed to the structural constraints imposed by the increasing $N_a$. The increase in the number of arms in an MSP molecule results in a more compact structure and thus higher segmental density than a linear BCP of the same molecular weight, which leads to $\mu \rightarrow 1/3$ \cite{Pakula1998, Rubin2007, Jack2017}. Consequently, this increased compactness causes MSPs to resemble \emph{soft dimeric colloids} in the melt \cite{Jack2017}, as illustrated by the far-right individual MSP images in Fig.~\ref{Fig2}. Furthermore, an increase in $N_a$ within an MSP molecule reduces the diffusion coefficient, $D$. However, the segregation current is multiplied by $N_b$, yielding an overall prefactor of $N_{b}^{1-\nu}$ for the growth law in the early stages of phase separation, i.e., before the topological constraint becomes relevant. However, at larger length scales, we expect a saturation into a lamellar structure analogous to BCP. In other words, the compatible MSP arms start forming local intramolecular clusters of $A$ and $B$ phases at the beginning of the growth process. As a result, MSPs with larger $N_a$ rapidly enhance the overall cluster size when they merge with other molecules or clusters despite their relatively slower diffusion rate. Note that the initial sequencing of $A$ and $B$ MSP arms does not affect the phase separation kinetics. Therefore, the segregation kinetics of MSP melt is governed by the delicate interplay of MSP molecule size leading to intramolecular domain formation and the diffusion coefficient \cite{Pakula1998, Rubin2007, Jack2017}. However, for simpler molecules like BCP, the primary factor influencing the average domain size is the diffusion coefficient \cite{Jack2017}.  

In Fig.~\ref{Fig3}(b), we depict the scaled correlation function, $C(r,t)$ vs. $r/L$, at $t=2.0\times 10^4$ for $N_a = 4$, $6$, $10$, $14$, $18$, and $22$, respectively. The zero crossing of $C(r,t)$ is indicated by the black horizontal line. The substantial overlap of $C(r,t)$ data for different $N_a$ values suggests that MSP melts with varying numbers of arms are part of the same dynamical universality class. The tail of the scaled $C(r,t)$ exhibits pronounced oscillatory behavior, indicating the formation of periodically ordered morphologies at late times. Moreover, the excellent dynamical scaling in $C(r, t)$ indicates that SMSPs asymptotically evolve into the same ordered lamellar structure.

We present the characteristic length scale, $L(t)$, as a function of time, $t$, in Fig.~\ref{Fig3}(c) on a logarithmic scale for $N_a = 4$, $6$, $10$, $14$, $18$, and $22$. The growth kinetics for $N_a = 4$ (red) is slightly faster than for $N_a = 6$ (blue). As previously discussed, the predominant factor influencing the average cluster size for smaller MSP molecules is diffusion rather than intramolecular cluster formation, which explains this observation. However, as $N_a$ increases symmetrically, local intramolecular domains begin to form. This results in a consistent increase in the domain growth rate from the onset despite the reduced diffusion. Thus, the characteristic average domain growth is dependent on $N_a$. Nonetheless, the MSP system exhibits diffusive growth, $L(t) \sim t^{1/3}$, for all cases at early times (indicated by a solid black line with a slope of $1/3$) before transitioning to saturation into a lamellar structure analogous to BCP melt at later times.

In Fig.~\ref{Fig3}(d), we analyze the scaling behavior of $L(t)$ vs. $t$ plots to study the universality of the characteristic length in segregating MSP melts. Using $N_a = 4$ as the reference, we scale each dataset as $L(t)$ vs. $t/t_0$, where $t_0$ is the time scaling factor associated with distinct $N_a$ values. The data collapses when we scale the $t$-axis by $t_0 \simeq \left(N_a / 4\right)^{-\theta}$ where $\theta$ is the exponent. The substantial overlap in the $L(t)$ vs. $t/t_0$ data suggests that MSP melts with varying $N_a$ belong to the same universality class. This observation indicates that SMSP melts with different $N_a$ values exhibit the same ordered morphologies in the asymptotic regime. The inset shows the variation of $t_0$ as a function of $N_a$ on a log-log plot, revealing a power-law decay with an exponent $\theta \simeq 1.0$, depicted by the black solid line. This relationship underscores the universal nature of the phase separation dynamics of SMSP melts across different $N_a$ values.

%%%%%%%%%%%%%%%%%%%%%%%%%%%
\subsection{Asymmetric MSP melt}
\begin{figure}[tb]
 \centering
 \includegraphics[width=0.7\textwidth]{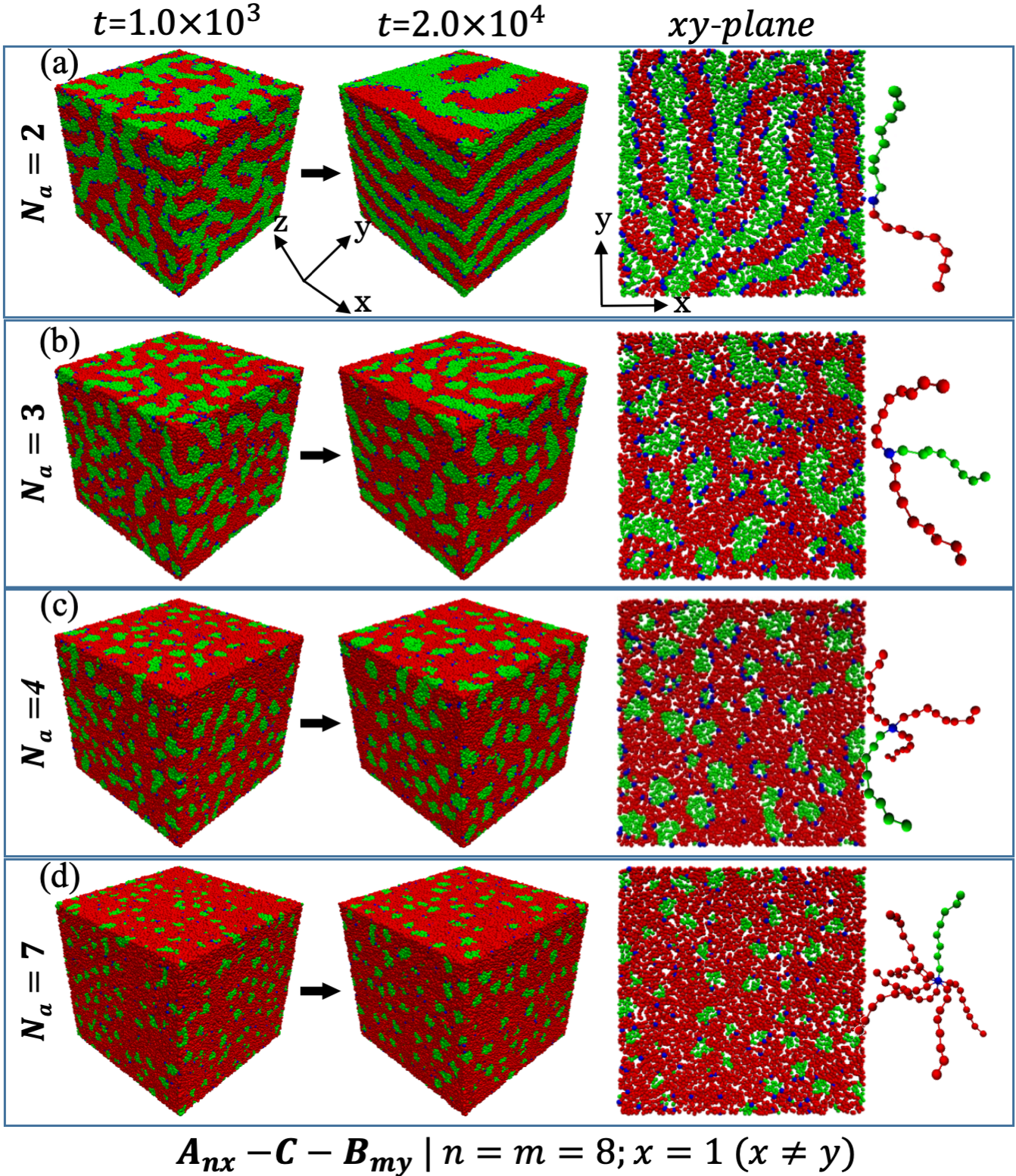}
 \caption{The first and second columns are $3d$ evolution snapshots of AMSP melts ($n=m$, $x\neq y$ with $x=1$) at $t=1.0\times10^3$ and $t=2.0\times10^4$ respectively, for $N_{a}=2$, $3$, $4$, and $7$. The third column displays the $2d$ cross-sectional plots at $t=2.0\times10^4$. On the extreme right, we display the microscopic images of individual AMSP molecules in the melt.} 
 \label{Fig4}
\end{figure}

In this section, we analyze the phase separation kinetics of AMSP melts of type $A_{nx}-C-B_{my}$ where $n = m$ and $x\neq y$ with only one $A$-type of arm, i.e., $x=1$. We study the segregating AMSP melt as a function of $y$, providing a thorough insight into the interplay between molecular architecture and emergent morphologies. The evolution snapshots for cases $N_a=2$, $3$, $4$, and $7$ are shown in Fig.~\ref{Fig4} at two distinct times, $t=1.0\times10^3$ (in first column) and $t=2.0\times10^4$ (in second column). To comprehend the late-time evolution structures, we showcase the $xy$ cross-section at $z=32$ of the $3d$ morphologies at $t=2.0\times10^4$ in the third column. The far-right images depict microscopic views of individual MSP molecules within the melt. 

The phase separation for the $N_{a}=2$ MSP case is similar to that of a linear symmetric BCP melt. The system undergoes \emph{spinodal decomposition} in the early stages and forms lamellar structures at later times. Over time, these systems exhibit microphase separation into periodic domains due to the balanced repulsive and attractive interactions between blocks, resulting in a stable lamellar morphology (see Fig.~\ref{Fig4}(a)). However, the lamellar ordering for $N_{a}=2$ is disrupted when there is asymmetry (off-criticality) in MSP molecules. In the MSP melt for $N_a=3$ (one green arm ($x=1$) and two red arms ($y=2$)), the system undergoes \emph{nucleation and growth} at early times and gradually microphase separates into a peanut-shaped morphology. As the off-criticality increases further, the AMSP system for $N_a \geq 4$ evolves into a spherical droplet morphology at later times, where spherical droplets of the $A$ phase (green) are observed within a continuous $B$ phase (red background). These visual representations illustrate how increasing asymmetry, quantified by the increase in $y$, significantly influences the structural organization within the MSP melts. The resulting morphologies resemble those seen in highly off-critical BCP melts undergoing phase separation.

\begin{figure}[tb]
\centering
  \includegraphics[width=0.7\textwidth]{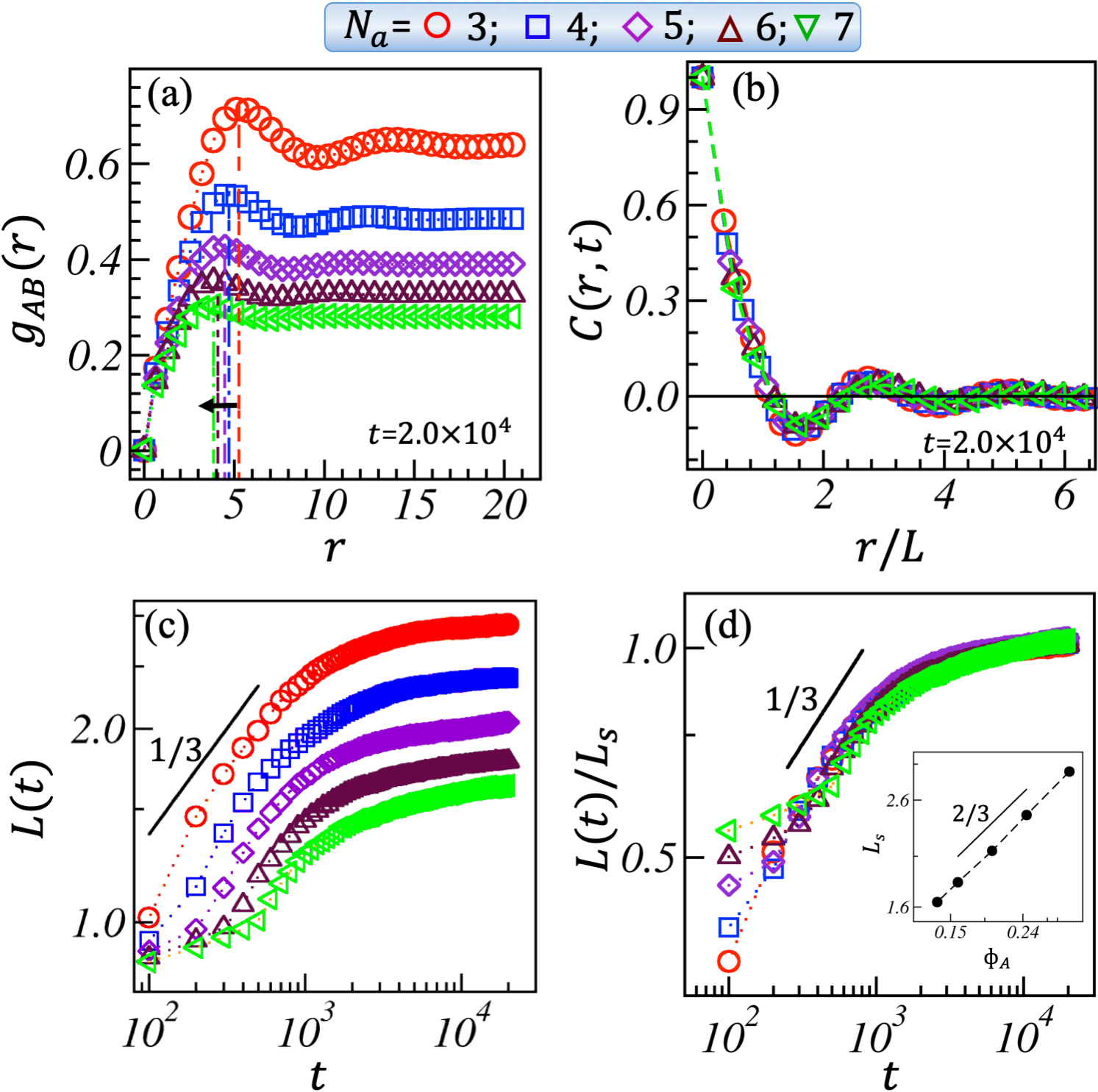}
  \caption{(a) RDF, $g_{AB}$ vs. $r$, at $t=2.0\times10^4$ for $N_{a}=3$ (red), $4$ (blue), $5$ (purple), $6$ (maroon), and $7$ (green). (b) Scaled correlation function, $C(r,t)$ vs. $r/L$, at $t=2.0\times10^4$. (c) Log-log plot of the characteristic length scale, $L(t)$ vs. $t$. A solid black line of slope $1/3$ displays the diffusive growth regime. (d) Scaling of the average domain size, $L(t)/L_S$ vs. $t$; $L_S$ is the saturation length. Inset shows the variation of $L_S$ as a function of $\phi_A$, volume fraction of $A$-phase.}
  \label{Fig5}
\end{figure}

To quantitatively analyze the system, we depict the RDF ($g_{AB}$ vs. $r$) at $t=2.0\times10^4$ in Fig.~\ref{Fig5}(a). We observe a gradual reduction in the amplitude of RDF peaks with increasing asymmetry ($y=2\rightarrow 6$ with $x=1$ fixed) of the MSP molecule. This indicates a tendency toward looser clustering of distinct phases. Given that the total bead density in the system is fixed, an increase in the number of $y$ arms (comprising $B$-type beads shown in red) leads to a relative decrease in the volume fraction of $A$-type beads. Consequently, the shift in peak positions towards lower $r$ values reflects a reduction in the cluster size of the $A$-phase. These results highlight the intricate impact of molecular asymmetry on the structural characteristics of the MSP system. 

We studied the statistical behavior of various morphologies by plotting the scaled correlation function, $C(r,t)$ vs. $r/L$, at a late time $t=2.0\times10^4$, as shown in Fig.~\ref{Fig5}(b). The scaling plot for $N_a = 3$ (red) slightly deviates from the scaling functions obtained for $N_a \geq 4$, exhibiting excellent data overlap. This suggests that AMSP morphologies with increasing $N_a$ can be classified within the same dynamic universality class, indicating similar dynamics and structural arrangement within the system. Additionally, we observe a jump in $C(r,t)$ data as $r/L \rightarrow 0$, which becomes more pronounced with increasing $N_a$. This further suggests that cluster interfaces are getting fuzzier with increasing $N_a$, confirming the observation of loose particle clustering. Notice the oscillatory behavior of the scaled $C(r,t)$ at the late time $t = 2.0 \times 10^4$, which indicates the evolution of nearly similar morphologies, specifically spherical droplet shapes in all cases.

In the current context of domain coarsening in a $3d$ AMSP system characterized by a saturation length $L_s$, the volume fraction of $A$-type beads, $\phi_A$, can be derived as follows:
\begin{equation}
\label{phiA1}
N_d \left(\dfrac{4\pi}{3}L_s^3\right)/L^3 \sim \phi_A,
\end{equation}
where, $\phi_A=x/(x+y)$, $N_d$ denotes the average number of domains, and $L$ is the fixed size of the simulation box. Thus, the above equation can be approximated to:
\begin{equation}
\label{phiA2}
N_dL_s^3 \sim \phi_A.
\end{equation}
Considering a constant total bead density $\rho=3$, an increase in $y$ results in a higher number of beads ($N_b$) per MSP molecule, reducing its diffusion coefficient and lowering the total number of MSP molecules in the system. With $x=1$ fixed, $\phi_A$ decreases as $y$ increases. The reduced diffusion rate of larger MSP molecules results in smaller cluster sizes at any given time, as illustrated by the green clusters in Fig. \ref{Fig4}(d). Therefore, $N_b\propto N_d$ as $y$ varies from $N_a=3$ to $7$, leading to the relationship $N_d \propto 1/\phi_A$. Consequently, from Eq. (\ref{phiA2}), we obtain:
\begin{equation}
\label{phiA3}
L_s \sim \phi_A^{2/3}.
\end{equation}

Subsequently, we plot the characteristic length scale, $L(t)$ vs. $t$, in Fig.~\ref{Fig5}(c) on a log-log scale. A clear trend emerges from these plots: increasing the asymmetry by increasing $y$ reduces the average domain size. Typical of phase-separating BCP melts, $L(t)$ initially follows a diffusive growth, $\phi \sim 1/3$, for all cases. Over time, the system crossovers into a saturation regime, validating the pinning of evolution morphologies. Figure~\ref{Fig5}(d) presents the scaling behavior of $L(t)$ vs. $t$ datasets for $N_a=3 \rightarrow 7$ to study the dynamic universality of the phase-separating AMSP system. We scale our data as $L(t)/L_s$ vs. $t$, where $L_s$ denotes the saturation length corresponding to each $N_a$ value. We observe a significant overlap in the scaled data for all cases, especially during the late growth times. However, there is a notable deviation in the early transient growth regimes. This discrepancy can be attributed to the limited range of $L(t)$ at early times normalized against a much broader range of $L_s$ at late times. These results indicate that the characteristic length scale of AMSP melts for different $N_a$ values follows the same dynamic universality. The inset shows the variation of the saturation length, $L_s$, with $\phi_A$ on a log-log scale, revealing a power-law relation expressed as $L_{s}\sim \phi_A^{\lambda}$, where the power-law exponent $\lambda \simeq 2/3$, as demonstrated in Eq. (\ref{phiA3}). Thus, our results highlight the diverse spectrum of dynamic behaviors of different AMSPs.
\begin{figure}[htb]
 \centering
 \includegraphics[width=0.45\textwidth]{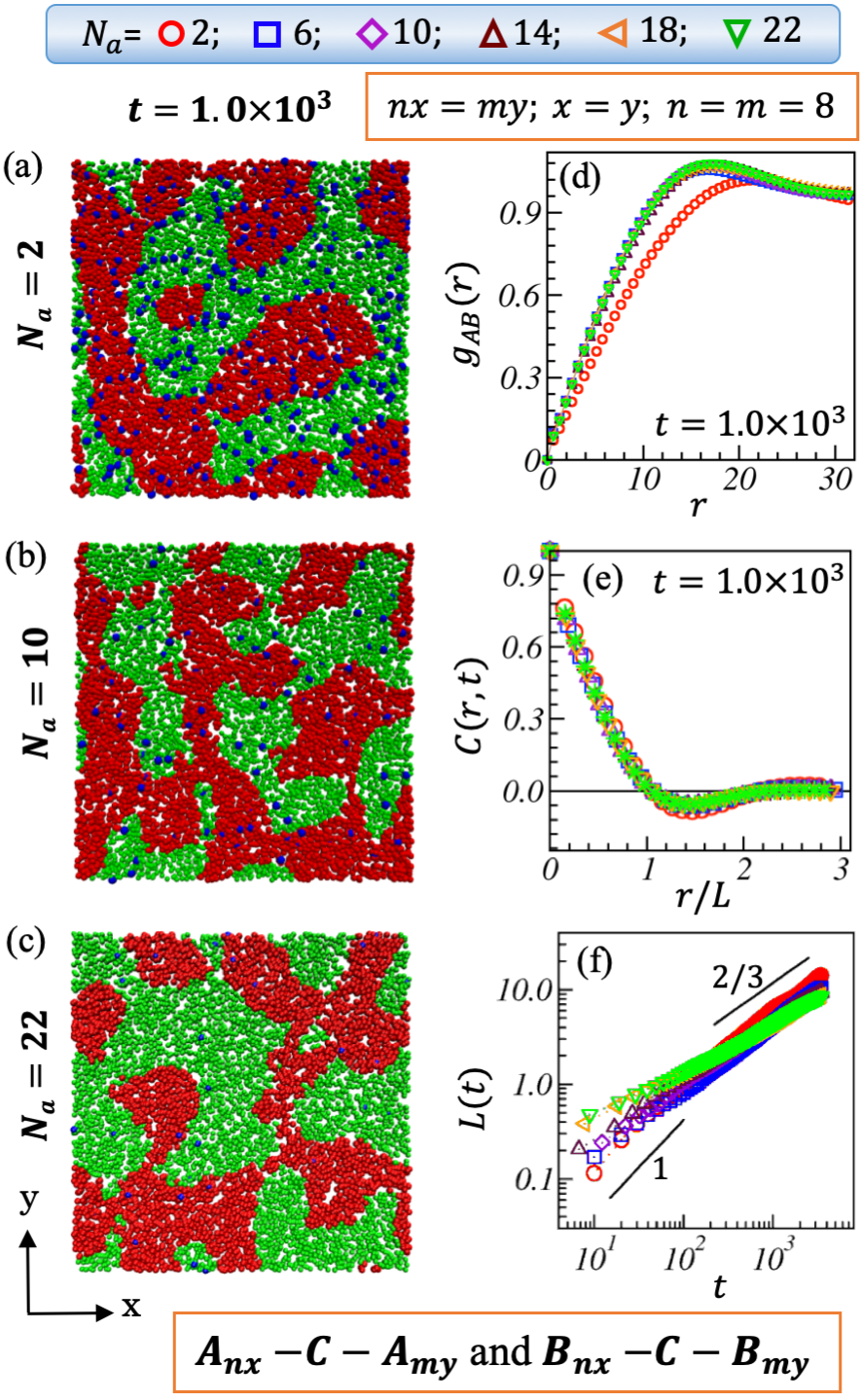}
 \caption{(a-c) $2d$ cross-section snapshots at $t=1.0\times 10^3$ for SMSP blends with $N_{a}=2$, $10$, and $22$. (d-f) RDFs ($g_{AB}$ vs. $r$), scaled correlation functions ($C(r,t)$ vs. $r/L$), and characteristic length scales ($L(t)$ vs. $t$) for $N_a=2$ (red), $6$ (blue), $10$ (purple), $14$ (maroon), $18$ (orange), and $22$ (green). The solid black lines with slopes $1$ and $2/3$ indicate the viscous and inertial hydrodynamic regimes, respectively.}
 \label{Fig6}
\end{figure}

%%%%%%%%%%%%%%%%%%%%%%%%%%%
\subsection{Phase separation in symmetric/asymmetric MSP blends}
In this section, we examine the evolution kinetics of binary MSP blends ($A_{nx}-C-A_{my}$ and $B_{nx}-C-B_{my}$) as depicted in Fig.~\ref{Fig6} for a different number of arms, $N_a$. First, we present the $2d$ snapshots ($xy$-plane at $z=32$) of the $3d$ macrophase-separated domains for the SMSP blends for $N_a=2$, $10$, and $22$ in Fig.~\ref{Fig6}(a-c)(see the first column) at a time, $t=1.0\times 10^3$. Since the number of $C$ beads (marked in blue) is much less than that of $A$ (red) and $B$ (green) beads, we mainly treat our systems as binary fluids. Note that $N_a=2$ implies a linear binary polymer blend system. To analyze these morphologies, we plot RDFs, $g_{AB}(r)$ vs. $r$, in Fig.~\ref{Fig6}(d) at $t=1.0\times 10^3$. For $N_a=2$, the RDF peak position is at a higher $r$, indicating a larger average cluster size. The overlap of RDF curves at late times for AMSP blends with a higher number of arms suggests a similar cluster size distribution for SMSP blends within our simulation timescale.

The scaled correlation functions, $C(r,t)$ vs. $r/L$ at $t=1.0\times 10^3$, shown in Fig.~\ref{Fig6}(e) for various $N_a$ values, exhibit excellent data overlap. This indicates that the morphological evolution of symmetric binary MSP blends, characterized by an increasing number of arms with the same molecular weight (see Fig.~\ref{Fig1}(c)), asymptotically falls within the same dynamical universality class. The consistent overlap of scaling functions underscores the robustness of phase separation dynamics in multicomponent SMSP blends, suggesting a common underlying mechanism. Notably, only a minor deviation in the scaled $C(r,t)$ is observed for $N_a=2$ at larger $r/L(t)$. Consequently, the phase separation behavior of SMSP blends conforms to the same dynamical universality class as that of linear polymer blends.
\begin{figure}[htb]
 \centering
 \includegraphics[width=0.45\textwidth]{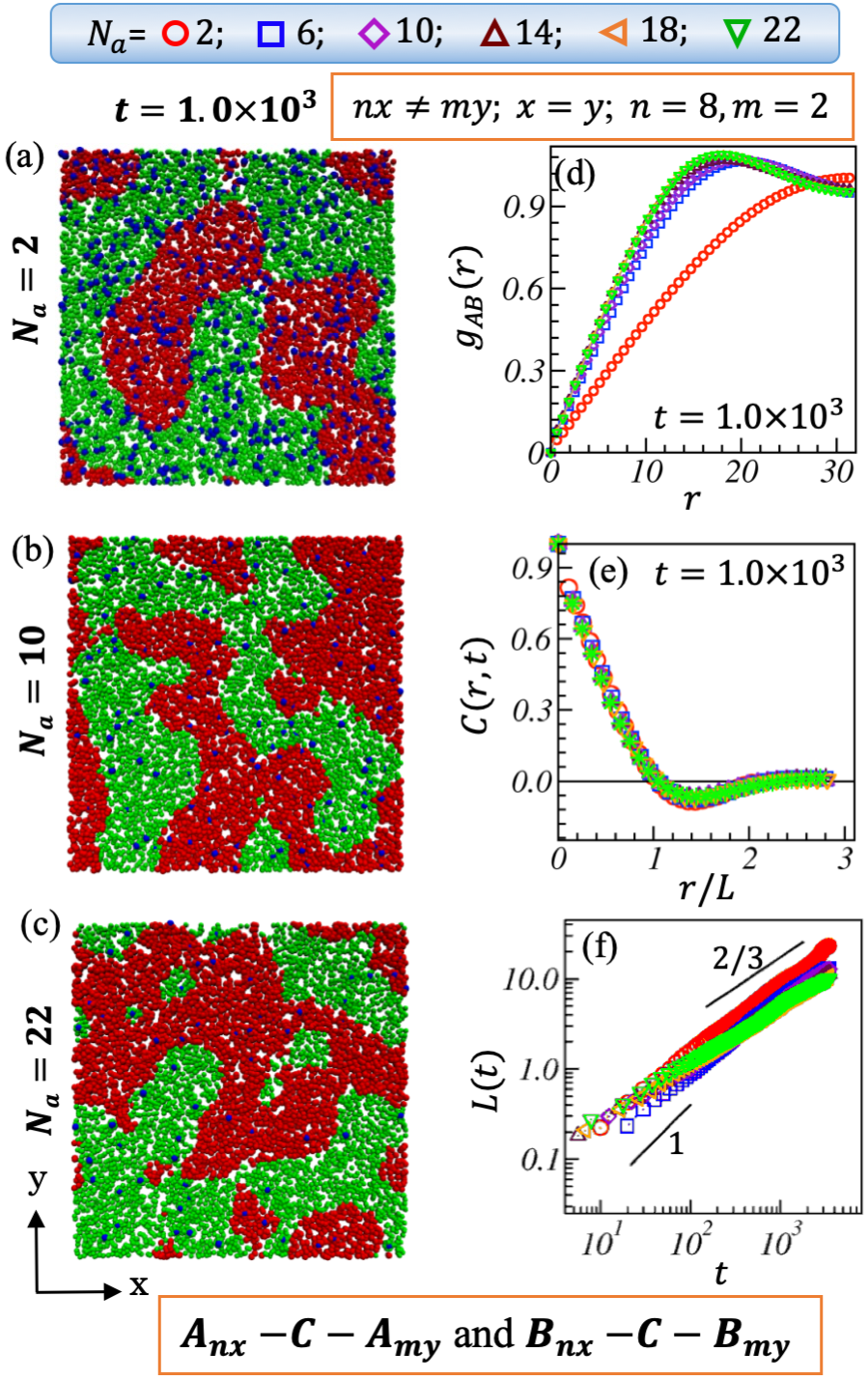}
 \caption{(a-c) $2d$ cross-section images at $t=1.0\times 10^3$ for AMSP blends with $N_{a}=2$, $10$, and $22$. (d-f) RDFs, $g_{AB}$ vs. $r$, scaled correlation functions, $C(r,t)$ vs. $r/L$, and characteristic length scales, $L(t)$ vs. $t$, for $N_a=2$ (red), $6$ (blue), $10$ (purple), $14$ (maroon), $18$ (orange), and $22$ (green). Solid black lines with slopes $1$ and $2/3$ indicate viscous and inertial hydrodynamic regimes.}
 \label{Fig7}
\end{figure}

We present the time evolution of the characteristic length scales, $L(t)$ vs. $t$, on a log-log scale in Fig.~\ref{Fig6}(f). For $N_a=2$, the growth exponent shows the expected transition from $\phi \sim 1$ (viscous hydrodynamic regime) at early times to $\phi \sim 2/3$ (inertial hydrodynamic regime) at later times, confirming the appropriateness of our chosen DPD interaction parameters. Black solid lines with slopes of 1 and 2/3 are included for reference. Note that the diffusive growth regime is very short-lived for macrophase-separating blends in DPD simulations \cite{singh2018role, singh2021photo}. Figure~\ref{Fig6}(f) indicates that as $N_a>2$, the average domain growth rate increases, while the growth exponent decreases consistently within the viscous hydrodynamic regime. This can be attributed to the reduction in the diffusion coefficient as molecular weight increases ($D\sim N_b^{-\nu}$). Concurrently, the transport of the average MSP mass ($R_g \sim N_b^{\mu}$) to the evolving domain is enhanced at a given time. Notably, SMSP blends with varying numbers of arms eventually reach the inertial hydrodynamic growth regime with the growth exponent of $\phi\sim 2/3$, indicating a nearly similar growth rate at late times as for the $N_a=2$ case, irrespective of the number of arms. This phenomenon can be understood by recognizing the fact that as the number of arms ($N_a$) increases, the corresponding decrease in $D$ is offset by the increase in average MSP mass transport to their respective domains during the late times.

Further, we present $2d$ cross-section images of macrophase-separated domains of AMSP blends for varying $N_a$ at $t=1.0\times 10^3$ in Figs.~\ref{Fig7}(a-c). Recall that the AMSP blend considered here has asymmetry in the molecular weight (degree of polymerization) of the arms, i.e., each MSP has an equal number of longer and shorter arms. We plot the RDF ($g_{AB}(r)$ vs. $r$) at $t=1.0\times 10^3$ in Fig.~\ref{Fig7}(d) to analyze the evolving morphologies for $N_a=2$, $6$, $10$, $14$, $18$, and $22$. Similar to the SMSP blend case, the RDF peak for $N_a=2$ occurs at a larger $r$. As $N_a$ increases, the RDF curves shift to lower $r$, indicating a smaller cluster size distribution. At higher $N_a$ values, such as $N_a=10$, $14$, $18$, and $22$, the RDF curves significantly overlap, suggesting an almost identical cluster size distribution asymptotically.

The dynamic scaling of correlation functions, $C(r,t)$ vs. $r/L$, at $t=1.0\times 10^3$ in Fig.~\ref{Fig7}(e) exhibits complete data overlap for different values of $N_a$. However, only a slight deviation is observed for $N_a=2$, particularly at larger $r/L$ values. Thus, akin to the SMSP blend case, the morphological evolution of binary AMSP blend, characterized by intra-arm molecular weight disparities, conforms to the same dynamical universality class as linear polymer blends in the late times. The length scale plots for the AMSP blends, depicted in Fig.~\ref{Fig7}(f), consistently align with a growth exponent of $\phi \sim 1$ with minimal deviation at early times. This implies that the increasing number of arms $N_a$ hardly influences the growth law exponents and the growth rate of phase-separating domains compared to SMSP blends at early times. Asymptotically, the length scale crosses over to an inertial hydrodynamic growth regime ($\phi \sim 2/3$) for all $N_a$ values. Similar to the SMSP cases, the length scale data for $N_a > 2$ collapses reasonably well, confirming a consistent average domain size growth rate. However, for $N_a =2$, the domain growth rate is slightly higher due to a larger diffusion coefficient than other $N_a$ values. Thus, the phase separation kinetics of AMSP blends mirror the SMSP blends, sharing the dynamical universality class with late-stage binary linear polymer blends. 

%%%%%%%%%%%%%%%%%%%%%%%%%%%
\section{Conclusion}
We utilized the dissipative particle dynamics (DPD) simulation technique to study the growth kinetics in miktoarm star polymer (MSP) melts and blends with different architectures. We noted interesting behavior in the microphase-separating symmetric MSP (SMSP) melts. For a given arm length (arm's molecular weight) of the MSP molecule, the growth rate of the average domain size decreased as the number of arms increased up to $N_a \simeq 6$. However, for $N_a > 6$, SMSPs exhibited effective local intra-molecular clustering. This clustering enhanced material transport to the evolving domains, leading to a higher growth rate despite the larger MSPs having a lower diffusion coefficient. Thus, we demonstrated a complex interplay between the diffusion of MSP molecules and their sizes, identifying a transition in the domain growth rate around $N_a \approx 6$. Moreover, we determined that SMSP melts with different numbers of arms belong to the same dynamical universality class.

During the early stages of evolution, the typical diffusive growth regime ($\phi \sim 1/3$) was observed, which asymptotically crossed over into a saturation regime, forming a lamellar structure analogous to block copolymer (BCP) melts at late times for all $N_a$ values of the SMSP melts studied. Nonetheless, we observed a significant overlap in the characteristic domain growth plot for SMSP melts when scaling the time as $t/t_0$, where $t_0$ is a scaling factor associated with distinct $N_a$ values. This further indicates that the segregation kinetics of SMSP melts with varying $N_a$ belong to the same dynamic universality class. The time scaling factor demonstrated a power-law decay, $t_0 \sim \left(N_a/4 \right)^{-\theta}$, with an exponent $\theta \simeq 1.0$ for the number of arms increases beyond $N_a \simeq 6$.

Given the architecture of AMSPs we studied, the melt system becomes highly off-critical when the number of one type of MSP arms increases. The evolution of AMSP melts yields a lamellar and peanut-shaped morphology for AMSPs with $N_a=2$ and $3$ arms. However, as $N_a$ exceeds $3$, spherical droplets emerge via nucleation and growth. The perfect overlap of dynamical scaling functions at late times confirms that AMSP melts with varying $N_a$ belong to the same dynamic universality class. Increasing asymmetry in AMSP melts reduces the growth rate of the average domain size. Similar to SMSP melts, the length scale shows typical diffusive growth ($\phi \sim 1/3$) at early times, which asymptotically saturates, indicating the pinning of morphology. The scaling of the length scale $L(t)$ with their corresponding saturation length $L_s$ as $L/L_s$ vs. $t$ depicts excellent data overlap at late times, with the expected deviation only in the early transient regimes. This further confirms the same dynamic universality class of AMSP melts with varying $N_a$. The variation of $L_s$ against the volume fraction of the lower component $\phi_A$ follows a power-law dependence: $L_s \sim \phi_A^{\lambda}$, where $\lambda \simeq 2/3$, corroborating our theoretical prediction.

Additionally, the SMSP and AMSP blends exhibit a similar dynamical universality class as linear polymer blends at late times. However, a slight deviation is observed at the early stages for AMSP blends. Early-time analysis of the length scale indicates that varying $N_a$ primarily affects the growth rate for SMSPs while having little effect on AMSPs. However, the growth rate stays nearly the same at late times for both blends. For SMSPs, the viscous hydrodynamic growth law exponent ($\phi \sim 1$) reduces with $N_a$, while AMSP melts are hardly affected. However, length scales tend towards inertial hydrodynamic growth asymptotically ($\phi \sim 2/3$). This suggests a balance between a decrease in the diffusion coefficient and enhanced mass transportation as $N_a$ increases. Overall, our results provide insights into the complex phase separation kinetics of MSP melts and blends, aiding in the design and manipulation of advanced polymeric systems.

\section*{Associated Content}
\subsection*{Data Availability Statement} Data will be made available on request.
\section*{Author Information}
\subsection*{Corresponding Authors}
\textbf{Sanjay Puri}- School of Physical Sciences, Jawaharlal Nehru University, New Delhi-110067, India.\\
\textbf{Awaneesh Singh}- Department of Physics, Indian Institute of Technology (BHU), Varanasi-221005, Uttar Pradesh, India; Email: awaneesh.phy@iitbhu.ac.in; orcid.org/0000-0002-8249-4520
\subsection*{Authors}
\textbf{Dorothy Gogoi}- School of Physical Sciences, Jawaharlal Nehru University, New Delhi-110067, India.
\textbf{Avinash Chauhan}- Department of Physics, Indian Institute of Technology (BHU), Varanasi-221005, Uttar Pradesh, India.
\subsection*{Author Contributions}
Dorothy Gogoi and Avinash Chauhan: Conceptualization, Methodology, Formal analysis and Investigation, Resources and Data curation, Writing-original draft, Visualization. Sanjay Puri and Awaneesh Singh: Conceptualization, Validation, Formal analysis and Investigation, Writing-review and editing, Supervision, Funding acquisition.
\subsection*{Note}
The authors declare no competing financial interest.
\section*{Acknowledgments}
D.G. acknowledges the HPC resources provided for computational support by Jawaharlal Nehru University, New Delhi. A.C. thanks IIT (BHU) for financial assistance. S.P. is grateful for the J.C. Bose fellowship from SERB, New Delhi, India. A.S. acknowledges the financial support from SERB under grant no. ECR/2017/002529 and CRG/2023/001311, New Delhi, India.
%%%%%%%%%%%%%%%%%%%%%%%%%%%%%%%%
%\footnotesize\bibliography{reference} 
\bibliography{reference}

\end{document}